\documentclass[a4paper,reqno,12pt,final]{amsart}
\usepackage{amsmath,amsfonts,amssymb,amsthm,amstext,eucal,bbold}
\usepackage{array}
\usepackage[latin1]{inputenc}
\DeclareMathOperator{\AdS}{AdS}
\DeclareMathOperator{\rank}{rank}
\DeclareMathOperator{\CW}{CW}

\DeclareMathOperator{\Mat}{Mat}

\DeclareMathOperator{\dvol}{dvol}

\renewcommand{\d}{\partial}
\newcommand{\1}{\mathbb{1}}
\newcommand{\Cl}{\mathrm{C}\ell}

\newcommand{\RR}{\mathbb{R}}

\newcommand{\EE}{\mathbb{E}}
\newcommand{\ZZ}{\mathbb{Z}}

\newcommand{\SU}{\mathrm{SU}}
\newcommand{\U}{\mathrm{U}}

\newcommand{\Spin}{\mathrm{Spin}}
\newcommand{\eA}{\mathcal{A}}
\newcommand{\eD}{\mathcal{D}}
\newcommand{\eF}{\mathcal{F}}

\newcommand{\eR}{\mathcal{R}}
\newcommand{\eS}{\mathcal{S}}

\newcommand{\fg}{\mathfrak{g}}
\newcommand{\fh}{\mathfrak{h}}
\newcommand{\fp}{\mathfrak{p}}

\newcommand{\fn}{\mathfrak{n}}
\newcommand{\fm}{\mathfrak{m}}
\newcommand{\half}{\tfrac{1}{2}}

\theoremstyle{plain}
\newtheorem{thm}{Theorem}

\newcommand{\MUNCH}[1]{\relax}
\allowdisplaybreaks[1]

\begin{document}

\title[Maximal supersymmetry and supergravity]{Maximally
  supersymmetric solutions of ten- and eleven-dimensional
  supergravities}
\author[Figueroa-O'Farrill]{José Figueroa-O'Farrill}
\address{School of Mathematics, University of Edinburgh, Scotland, UK}
\email{j.m.figueroa@ed.ac.uk}
\author[Papadopoulos]{George Papadopoulos}
\address{Department of Mathematics, King's College, London, England, UK}
\email{gpapas@mth.kcl.ac.uk}
\thanks{EMPG-02-16}
\date{\today}
\begin{abstract}
  We classify (up to local isometry) the maximally supersymmetric
  solutions of the eleven- and ten-dimensional supergravity theories.
  We find that the AdS solutions, the Hpp-waves and the flat space
  solutions exhaust them.
\end{abstract}

\maketitle

\tableofcontents

\section{Introduction and main results}

The investigation of string solitons, such as branes, has ushered in
an era of rapid progress in our understanding of nonperturbative
string theory.  At low energies these solitons are described by (often
supersymmetric) solutions of the corresponding supergravity theory.
Conversely, it is believed that these supergravity solutions can be
lifted to solutions of the full string theory equations of motion,
including all $\alpha'$ corrections.  Equally relevant are the
supersymmetric solutions of eleven-dimensional supergravity, as this
is the low-energy limit of M-theory, itself the strong coupling limit
of type IIA string theory.

Despite the huge catalogue of supersymmetric solutions of supergravity
theories, we are still far from possessing an overall picture of the
moduli space of such solutions and in fact until recently some basic
questions remained unanswered.

The most important invariant of a supergravity background is the
amount of supersymmetry that it preserves, usually labeled as a
fraction, traditionally denoted $\nu$, of the supersymmetry of the
vacuum.  In eleven-dimensional and type II supergravity theories, this
is a fraction taking values in the set
$\{0,\frac1{32},\tfrac1{16},\dots,\tfrac{31}{32},1\}$.  At the time of
writing it is now known whether all such fractions can in fact occur,
although some progress has been made recently in the construction of
backgrounds possessing hitherto unseen fractions $\half < \nu < 1$.

This fraction is defined as follows.  We are only concerned with
solutions of the equations of motion where the fermionic fields are
set to zero.  Such a solution is supersymmetric if it is left
invariant under some nontrivial supersymmetry transformations.  These
transformations are parametrised by spinors and, since the fermions
have been put to zero, the only nontrivial transformations are those
of the fermions themselves.  The supersymmetric variation of the
gravitino $\Psi_M$ defines a covariant derivative $\eD$ which is
induced from a connection on the bundle of spinors:
\begin{equation}
  \label{eq:KS1}
  \delta_\varepsilon \Psi_M = \eD_M \varepsilon~.
\end{equation}
The tensor which measures the deviation of $\eD$ from the spin
connection $\nabla$ depends algebraically on the bosonic fields of the
theory.  The other fermionic fields (if any) give rise to algebraic
equations of the form, say,
\begin{equation}
  \label{eq:KS0}
  \delta_\varepsilon \psi = \eA \varepsilon~,
\end{equation}
where $\eA$ is algebraic (i.e., zeroth order differential operator),
itself depending on the bosonic fields of the supergravity theory.  A
(real) spinor $\varepsilon$ is called a \emph{Killing spinor} if it
obeys the above equations, keeping in mind that the second equation
may not arise if there are no other fermionic fields beside the
gravitino---this happens for example, in eleven-dimensional
supergravity.

Because the above equations are linear, the Killing spinors form a
vector space whose dimension is at most the rank of the spinor bundle
$\eS$ of which the supersymmetry parameters $\varepsilon$ are
sections, which for the theories under consideration will be either
$32$ or $16$.  The reason is that equation \eqref{eq:KS1} says that a
Killing spinor is covariant constant with respect to the connection
$\eD$ and hence parallel transport uniquely defines a Killing spinor
on all points of the spacetime from its value at any given point.

Finally, the fraction $\nu$ is defined as the following ratio
\begin{equation*}
  \nu = \frac{\dim\{\text{Killing spinors}\}}{\rank \eS}~.
\end{equation*}

Equation \eqref{eq:KS1} allows us to define another invariant of the
solution which refines the fraction $\nu$, namely the holonomy
representation of the connection $\eD$.  This is a refinement of the
fraction $\nu$ because one can recover $\nu$ from the dimension of the
subspace of invariants in the spinor representation (subject perhaps
to the additional algebraic equations), and because different holonomy
representations actually give rise to the same fraction.  If we turn
off all fields but the metric, so that we consider a purely
gravitational solution, $\eD$ coincides with the spin connection
$\nabla$ whose holonomy group is contained in (the spin cover of) the
Lorentz group and, in the case of a supersymmetric background, must
also be contained in the isotropy group of a nonzero spinor.  As shown
in \cite{JMWaves,Bryant-ricciflat} for the case of eleven-dimensional
supergravity, this can either be $\SU(5)$ or $\RR^9 \rtimes \Spin(7)$.
In the former case this leads to static spacetimes generalising the
Kaluza--Klein monopole \cite{KKmonopoleS,KKmonopoleGP,HKMKK}, whereas
the latter case corresponds to generalisations of the purely
gravitational pp-wave \cite{MWave} which involve lorentzian holonomy
groups acting reducibly yet indecomposably on Minkowski spacetime.  In
ten dimensions the situation is even simpler and the isotropy group of
a chiral spinor must be contained in $\RR^8 \rtimes \Spin(7)$
\cite{Bryant-ricciflat}.  When we turn on the other fields in the
background, the analysis of the connection $\eD$ is complicated by the
facts that $\eD$ is not induced from a connection on the frame bundle
and that its holonomy is generic.  A holonomy analysis along the lines
advocated in \cite{JMWaves} has not yet been performed for
supersymmetric solutions with flux.  However much progress has been
done in special cases \cite{PapaRotated, PTConifolds} leading to the
no-go theorem of \cite{IPNoGo} for compactifications with torsion.

In this paper we take a first step in this direction by classifying,
up to local isometry, all those solutions of eleven and
ten-dimensional supergravity theories for which the (restricted)
holonomy of $\eD$ is trivial; in other words, we classify the
maximally supersymmetric solutions or \emph{vacua}.

With the exception of the massive IIA theory which, as we will see
below, has no maximally supersymmetric background, every other
supergravity theory in ten and eleven dimensions has a ``trivial''
vacuum in which the metric is flat and there are no fluxes.  In
addition, it has been known for some time that both eleven-dimensional
and type IIB supergravities have vacua of the form $\AdS_{p+2} \times
S^{D-p-2}$ \cite{FreundRubin,AdS7S4,SchwarzIIB} for $(D,p)\in
\{(11,2),(11,5),(10,3)\}$.

It is now well known that given any solution of a supergravity theory,
its plane wave limit \cite{PenrosePlaneWave, GuevenPlaneWave} yields
another solution which, as proven in \cite{Limits}, preserves at least
as much supersymmetry as the original solution.  It follows that the
plane wave limit of a maximally supersymmetric solution will also be
maximally supersymmetric.  Indeed it was shown in \cite{ShortLimits}
that taking a plane wave limit of the $\AdS_4 \times S^7$ and $\AdS_7
\times S^4$ vacua of eleven-dimensional supergravity one recovers the
maximally supersymmetric plane wave discovered by Kowalski-Glikman
\cite{KG,CKG} and re-discovered in \cite{FOPflux} where it was
identified as an \emph{Hpp-wave}; that is, a plane wave whose geometry
is that of a lorentzian symmetric space \cite{CahenWallach} and whose
fluxes are homogeneous.  Similarly, it was also shown in
\cite{ShortLimits}, that a plane wave limit of the $\AdS_5 \times S^5$
vacuum of type IIB supergravity yields the recently discovered
maximally supersymmetric Hpp-wave solution \cite{NewIIB}.
Furthermore, it was also shown in \cite{Limits} that other plane wave
limits yield either the flat vacuum or again the known maximally
supersymmetric Hpp-waves, so that no further vacua are obtained in
this way.  In this paper we will prove that there are no further
maximally supersymmetric solutions in any of the eleven- and
ten-dimensional supergravity theories.\footnote{The AdS solution of
  type I supergravity found in \cite{KGTypeI} is erroneously claimed
  to be maximally supersymmetric when in fact it preserves only
  one-half of the supersymmetry.}  Our proof consists of a systematic
investigation of the conditions which follow from demanding that the
curvature of the connection $\eD$ vanishes.  This analysis will in
fact allow us to (re)derive the existence of all the known maximally
supersymmetric solutions.  In particular we will derive the
Freund--Rubin ansatz from maximal supersymmetry.  It is perhaps
remarkable that it is actually possible to characterise these
conditions geometrically and hence determine exactly all the
solutions.

In addition to being a first step in the classification programme of
supersymmetric backgrounds, the study of vacua is physically
interesting since every vacuum defines a different stable sector of
the theory in which to study excitations, both perturbative and
solitonic.  This has been extensively studied for the flat vacuum and
to some extent for the AdS vacua, but in principle all vacua are to be
treated on the same footing.  Moreover the existence of the Hpp-wave
vacua and in particular their interpretation as plane wave limits, has
led to new progress on the AdS/CFT correspondence \cite{AdSCFTReview}
and in particular has led to a gauge-theoretic derivation
\cite{MaldaPL} of the spectrum of the IIB superstring on flat space
and on the Hpp-wave, which can be quantized exactly \cite{MetsaevIIB,
MetTseIIB} in the light-cone gauge.

It is natural to wonder whether it is possible to extend this
programme to solutions which preserve less than maximal supersymmetry,
as there are several classes of such solutions that may have
applications in string theory.  For example, strings in Hpp-waves can
be quantised exactly in the light-cone gauge and give rise to free
massive theories in two dimensions\cite{CLPppM}.  More generally,
strings on a larger class of pp-waves give rise to interacting massive
theories in two dimensions \cite{MaldacenaMaoz}.  The Hpp-wave ansatz
has also proven useful in constructing solutions with ``exotic''
fractions of supersymmetry: although the generic Hpp-wave solution
preserves one half of the supersymmetry, there are points in the
moduli space of Hpp-waves besides the one corresponding to the
maximally supersymmetric solution, where $\half < \nu < 1$
\cite{Michelson,CLPppM, ChrisJerome}.  Another interesting class of
solutions are those that preserve half of the supersymmetry and so
include all elementary brane solutions as well as certain
non-threshold bound states and other solutions in their U-duality
orbit.  The catalogue of such solutions is certainly large and a
classification is not known, but we believe that the systematic
approach advocated in this paper will prove useful in providing, if
not a classification, at least a geometric characterisation of such
solutions.

Our approach to the determination of the maximally supersymmetric
solutions is based on the following strategy.  Maximal supersymmetry
implies the flatness of the connection $\eD$, which becomes an
algebraic equation
\begin{equation}
  \label{eq:flatness}
  [\eD_M,\eD_N] = \eR_{MN} = 0~,
\end{equation}
which is supplemented (in some cases) by the algebraic equation
$\eA = 0$, derived from \eqref{eq:KS0}.  The operators $\eR_{MN}$ and
$\eA$ are endomorphisms of the spinor representation which can be
expanded in terms of antisymmetric products of gamma matrices of the
corresponding Clifford algebra.  Due to linear independence, the
coefficients of each antisymmetric product of gamma matrices must
vanish separately.  This gives a set of algebraic equations which we
will analyse and solve yielding the results we summarise below.

Before we outline the results of the paper, we need to introduce some
notation.  Throughout this paper we will use the notation $\CW_D(A)$
to denote the $D$-dimensional lorentzian symmetric space with metric
\begin{equation}
  \label{eq:CWmetric}
  g = 2 dx^+ dx^- + \left(\sum_{i,j=1}^{D-2} A_{ij} x^i x^j\right)
  (dx^-)^2 + \sum_{i=1}^{D-2} \left(dx^i\right)^2~,
\end{equation}
where $A = [A_{ij}]$ is a constant symmetric matrix.  More details on
these spaces can be found in \cite{CahenWallach,FOPflux} and a
concrete isometric embedding in $\EE^{2,D}$ can be found in
\cite{Limits}.   Similarly we will use the notation $\AdS_D(R)$ and
$S^D(R)$ to denote the $D$-dimensional anti~de~Sitter spacetime and
the $D$-dimensional sphere, respectively, where $R$ stands for the
value of the scalar curvature.

Let us now describe the results of this paper.  In
Section~\ref{sec:eleven} we will identify the vacua of
eleven-dimensional supergravity up to local isometry and prove the
following:

\begin{thm}
  \label{th:d=11}
  Let $(M,g,F_4)$ be a maximally supersymmetric solution of
  eleven-dimensional supergravity.  Then it is locally isometric to
  one of the following:
  \begin{itemize}
  \item $\AdS_7(-7R) \times S^4(8R)$ and $F = \sqrt{6 R} \dvol(S^4)$,
    where $R>0$ is the constant scalar curvature of $M$;
  \item $\AdS_4(8R) \times S^7(-7R)$ and $F = \sqrt{-6 R}
    \dvol(\AdS_4)$, where $R<0$ is again the constant scalar curvature
    of $M$; or
  \item $\CW_{11}(A)$ with $A=-\frac{\mu^2}{36}\, {\rm
      diag}(4,4,4,1,1,1,1,1,1)$ and\\
    $F = \mu\, dx^- \wedge dx^1\wedge dx^2 \wedge dx^3$.  One must
    distinguish between two cases:
    \begin{itemize}
    \item $\mu=0$: which recovers the flat space solution $\EE^{1,10}$
          with $F=0$; and
    \item $\mu\neq 0$: all these are isometric and describe an
          Hpp-wave.
    \end{itemize}
  \end{itemize}
\end{thm}

As a corollary we will determine the vacua of type IIA supergravity
and prove the following:

\begin{thm}
  \label{th:IIA}
  Any maximally supersymmetric solution of type IIA supergravity is
  locally isometric to $\EE^{1,9}$ with zero fluxes and constant
  dilaton.
\end{thm}

Theorems~\ref{th:d=11} and \ref{th:IIA} were announced in
\cite{Bonn}.

In Section~\ref{sec:IIB} we will determine the maximally supersymmetric
solutions of ten-dimensional type IIB supergravity and prove the
following result:

\begin{thm}
  \label{th:IIB}
  Let $(M,g,F_5^+,...)$ be a maximally supersymmetric solution of
  ten-dimensional type IIB supergravity.  Then it has constant axion
  and dilaton (normalised to 0 in the formulas below), all fluxes
  vanish except for the one corresponding to the self-dual five-form,
  and is locally isometric to one of the following:
  \begin{itemize}
  \item $\AdS_5(-R) \times S^5(R)$ and $F = 2 \sqrt{\frac{R}5}
    \left(\dvol(\AdS_5) + \dvol(S^5)\right)$, where $\pm R$ are the
    scalar curvatures of $\AdS_5$ and $S^5$, respectively; or
  \item $\CW_{10}(A)$ with $A=-\mu^2 \1$ and $F = \half \mu\, dx^-
    \wedge (dx^1\wedge dx^2 \wedge dx^3 \wedge dx^4 + dx^5\wedge dx^6
    \wedge dx^7 \wedge dx^8)$.  Again one must distinguish between two
    cases:
    \begin{itemize}
    \item $\mu=0$: which yields the flat space solution $\EE^{1,9}$
      with zero fluxes; and \item $\mu\neq 0$: all these are isometric
      and describe an Hpp-wave.
    \end{itemize}
  \end{itemize}
\end{thm}

The proof of this theorem rests on a conjectured Plücker-style formula
for orthogonal planes which we have verified for the case at hand, and
which we believe to hold in more generality.  The more general
conjecture as well as its verification in some cases is presented in a
separate article \cite{FOPPluecker}.

Finally in Section~\ref{sec:other} we will prove the following two
results on the remaining ten-dimensional supergravities:

\begin{thm}
  \label{th:other}
  Any maximally supersymmetric solution of type I or
  heterotic ten-dimensional supergravity is locally isometric to flat
  space with zero fluxes and constant dilaton.
\end{thm}

\begin{thm}
  \label{th:massive}
  Massive IIA supergravity has no maximally supersymmetric solutions.
\end{thm}

As a corollary of the determination of the maximally supersymmetric
solutions of ten- and eleven-dimensional supergravity theories, we can
determine all the maximally supersymmetric solutions of lower
dimensional supergravity theories which are obtained as toroidal
reductions from them, or more generally as quotients by a group
action.  In fact, it is not hard to show, using the methods in
\cite{FOPflux}, that the only vacua for these theories are flat with
vanishing fluxes and are obtained by quotienting the flat vacua in ten
and eleven dimensions by a translation subgroup of the isometries.

\section{Maximal supersymmetry in eleven-dimensional supergravity}
\label{sec:eleven}

\subsection{Eleven-dimensional supergravity}

The bosonic part of the action of eleven-dimensional supergravity
\cite{Nahm,CJS} is
\begin{equation}
  \label{eq:lag}
  \int_M \left( \half R \dvol - \tfrac14 F\wedge\star F + \tfrac1{12}
    F \wedge F \wedge A \right)~,
\end{equation}
where $F=dA$ is the four-form field strength, $R$ is the scalar
curvature of the metric $g$ and $\dvol$ is the (signed) volume element
\begin{equation*}
  \dvol := \sqrt{|g|}\, dx^0 \wedge dx^1 \wedge \cdots \wedge
  dx^{10}~.
\end{equation*}

The associated field equations of this action are as follows:
\begin{equation}
  \label{eq:ELexplicit}
  \begin{aligned}[m]
    d \star F &= - \half F \wedge F\\
    R_{MN} - \half g_{MN} R &= \half F^2_{MN} - \tfrac14 g_{MN} F^2~,
  \end{aligned}
\end{equation}
where we have defined the partial contractions
\begin{equation*}
  F^2_{MN} = \tfrac16 F_{MPQR} F_N{}^{PQR} \qquad\text{and}\qquad
  F^2 = \tfrac1{24} F_{MNPQ} F^{MNPQ}~,
\end{equation*}
and where
\begin{equation*}
  R_{MN} = R_{MPN}{}^P \qquad\text{and}\qquad
  [\nabla_M, \nabla_P]X^N=R_{MP}{}^N{}_Q X^Q~,
\end{equation*}
with $\nabla$ the Levi-Civita connection of $g$, $R$ the curvature of
$\nabla$ and $M,N,\dots=0,\dots,10$.  Notice that taking the trace of
the Einstein-type equation we obtain that
\begin{equation*}
  R = \tfrac16 F^2~.
\end{equation*}

In order to discuss supersymmetric solutions we have to say a word
about spinors.  We will be working with the Clifford algebra
\begin{equation}
  \label{eq:cliff}
  \Gamma_A \Gamma_B + \Gamma_B \Gamma_A = + 2 \eta_{AB} \1~,
\end{equation}
with $\eta$ the \emph{mostly plus} metric and the indices
$A,B=0,\dots, 10$ being frame indices.  In the standard notation (see,
e.g., \cite{LM}) this defines the Clifford algebra $\Cl(1,10)$ which
is isomorphic to $\Mat(32,\RR) \oplus \Mat(32,\RR)$ and so it has two
irreducible representations $S^\pm$ isomorphic to $\RR^{32}$.  These
are distinguished by the action of the volume element $\Gamma^{12}=
\pm \1$, respectively.  Of course, both $S^\pm$ are isomorphic, as
representations of $\Spin(1,10)\subset \Cl(1,10)$, to the unique
spinor representation $S$ of $\Spin(1,10)$.  In our conventions, the
gravitino belongs to $S^-$.

The supersymmetric variation of the gravitino defines the so-called
supercovariant connection
\begin{equation}
  \label{eq:gconnexplicit}
  \eD_M = \nabla_M - \tfrac1{288}\left( \Gamma^{PQRS}{}_M + 8
  \Gamma^{PQR}\delta^S_M \right) F_{PQRS}~,
\end{equation}
where the spin connection is given by
\begin{equation}
  \label{eq:spinconnexplicit}
  \nabla_M = \d_M + \tfrac14 \omega_M{}^{AB} \Gamma_{AB}~,
\end{equation}
where we have used the same symbol $\nabla$ to denote the Levi-Civita
connection of $g$ and the associated spin connection.

The supercovariant connection $\eD$ is a connection of
the bundle of spinors associated to the Clifford representation
$S^-$, which is \emph{not} induced from the frame bundle because of
the term proportional to the four-form $F$.
 
A solution of eleven-dimensional supergravity is supersymmetric if
it admits nonzero \emph{Killing spinors} $\epsilon$ defined by
\begin{equation}
  \label{eq:kseqn}
  \eD_M\epsilon=0~.
\end{equation}
The number of supersymmetries preserved by a solution is the maximum
number of linearly independent Killing spinors.  A solution is
maximally supersymmetric if it has thirty-two linearly independent
Killing spinors. This is equivalent to the connection $\eD$ having
trivial holonomy:
\begin{equation}
\label{eq:holcon}
  {\rm Hol}(\eD)= 1~.
\end{equation}
Iterating equation \eqref{eq:kseqn} leads to the integrability condition
\begin{equation}
  [\eD_M, \eD_N]\epsilon= \eR_{MN}\epsilon=0~,
\end{equation}
where $\eR_{MN}=[\eD_M, \eD_N]$ is the curvature of the supercovariant
connection.  A necessary condition for maximal supersymmetry is the
zero curvature condition:
\begin{equation}
  \label{eq:zerocurv}
  \eR_{MN}=0~.
\end{equation}
This equation can be expanded in a basis of the Clifford algebra
(modulo the centre), which is given by the skew-symmetric products of
gamma-matrices.  The requirement that $\eR=0$ implies that every
component of $\eR$ in this basis should vanish.

The zero curvature condition is equivalent to the triviality of the
\emph{restricted} holonomy of the supercovariant connection: that is,
the holonomy around contractible loops.  Hence on a simply-connected
spacetime, the zero curvature condition implies the holonomy condition
\eqref{eq:holcon}.  For a non-simply connected spacetime, the
integrability condition \eqref{eq:zerocurv} is necessary but not
sufficient.  In addition one must ensure that the Killing spinors are
either periodic or antiperiodic along noncontractible loops.  In this
paper we will focus on local solutions, and this distinction will not
arise.

These necessary and sufficient conditions for the existence of a
maximally supersymmetric spacetime can be easily extended to other
supergravities.  In the ten-dimensional supergravities treated in this
paper, the Killing spinor equation \eqref{eq:kseqn} is supplemented by
algebraic conditions coming from the supersymmetry variation of
fermionic fields other than the gravitino which are present in the
theory.  Therefore to find the solutions with maximal supersymmetry,
these algebraic Killing spinor equations have to be imposed in
addition to the vanishing of the curvature of the supercovariant
connection.  The additional algebraic Killing spinor equations do not
alter the above discussion of the existence of Killing spinors in
non-simply connected spacetimes.

\subsection{The zero curvature equations}

In this section we will study the equations which arise from demanding
the vanishing of the curvature $\eR$ of the supercovariant connection
$\eD$.  After some computation we find that (cf. \cite{BEdWN})
\begin{multline}
  \label{eq:pseudocurv}
  \eR_{MN}= \tfrac14 R_{MN}{}_{AB}  \Gamma^{AB} - \frac{2}{(288)^2}
  F_{M_1\dots M_4} F_{N_1\dots N_4}
  \epsilon_{MN}{}^{M_1\dots M_4 N_1\dots N_4}{}_L \Gamma^L\\
  + \tfrac{48}{(288)^2}\big[4 F_{MLPQ} F^{LPQ}{}_{M_1} \Gamma^{M_1}{}_N-
  4 F_{NLPQ} F^{LPQ}{}_{M_1}\Gamma^{M_1}{}_M \\
  -36 F_{LPMM_1} F^{LP}{}_{NN_1} \Gamma^{M_1N_1}
  +F_{L_1\dots L_4} F^{L_1\dots L_4} \Gamma_{MN}\big]\\
  +\tfrac{1}{36} \big[\nabla_M F_{NM_1M_2M_3}-\nabla_NF_{MM_1M_2M_3}\big]
  \Gamma^{M_1M_2M_3}
  \\
  +\tfrac{8}{(288)^2 3}\big[F_{M_1\dots M_4} F_{N_1\dots N_3N}
  \epsilon_M{}^{M_1\dots M_4N_1\dots N_3}{}_{L_1\dots L_3}
  - (N\leftrightarrow M)\big] \Gamma^{L_1L_2L_3}
  \\
  -\tfrac{1}{ 432}\big[4 F_{LM_1\dots M_3} F_{MN}^L{}_{N_1}
  \Gamma^{M_1\dots M_3N_1} \\
  +3 F_{LPM_1M_2} F^{LPN_1}{}_N \Gamma^{M_1M_2}{}_{M N_1}-3 F_{LPM_1M_2}
  F^{PLN_1}{}_M\Gamma^{M_1M_2}{}_{NN_1}\big]\\
  -\tfrac{1}{288}\big[\nabla_MF_{N_1\dots N_4} \Gamma^{N_1\dots N_4}{}_N-
  (N \leftrightarrow M)\big]
  \\
  +\tfrac{1}{(72)^2 5!} \big[-6 F_{MM_1\dots M_3} F_{NN_1\dots N_3}
  \epsilon^{M_1\dots M_3 N_1\dots N_3}{}_{L_1\dots L_5}\\
  - 6 F_{MPM_1M_2} F^P_{N_1\dots N_3}
  \epsilon_N{}^{M_1M_2N_1\dots N_3}{}_{L_1\dots L_5}
  \\
  +6 F_{NPM_1M_2} F^P_{N_1\dots N_3}
  \epsilon_M{}^{M_1M_2N_1\dots N_3}{}_{L_1\dots L_5}
  \\
  +9F_{PQM_1M_2} F^{PQ}{}_{N_1N_2}
  \epsilon_{MN}{}^{M_1M_2N_1N_2}{}_{L_1\dots L_5}\big]
  \Gamma^{L_1\dots L_5}~,
\end{multline}
where we have used that
\begin{equation}
  \Gamma^{A_1\dots A_{2k}}=\tfrac{(-1)^k}{ (11-2k)!} \epsilon^{A_1\dots
  A_{2k}}{}_{B_1\dots B_{11-2k}} \Gamma^{B_1\dots B_{11-2k}}~.
\end{equation}
Maximal supersymmetry demands that the coefficient of every term in
the above expansion of $\eR$ in skew-symmetric products of the
gamma-matrices should vanish.  This gives an over-determined system of
equations which we now analyse in turn.

\subsubsection{Terms linear in $\Gamma$}

The vanishing of the term in the curvature of the supercovariant
connection linear in the $\Gamma$-matrices implies that
\begin{equation}
  \label{eq:ling}
  F\wedge F=0~.
\end{equation}
Taking the inner product of this condition with respect to a vector
field $X$, we find that
\begin{equation}
  \label{eq:innerp}
  \iota_X F\wedge F=0~.
\end{equation}

\subsubsection{Terms quadratic  in $\Gamma$}

The vanishing of the term in the curvature of the supercovariant
connection quadratic in the $\Gamma$-matrices implies the following
equation:
\begin{multline}
  \label{eq:quaqua}
  R_{MNPQ} +
  \tfrac1{36}\left(F^2_{NP} g_{MQ} - F^2_{MP} g_{NQ} - F^2_{NQ} g_{MP}
    + F^2_{MQ} g_{NP}\right)\\
  - \tfrac1{12} \left(F^2_{MPNQ} - F^2_{MQNP}\right) + \tfrac1{36} F^2
    \left( g_{MP} g_{NQ} - g_{MQ} g_{NP} \right) = 0~,
\end{multline}
where we have introduced the partial contraction
\begin{equation*}
  F^2_{MNPQ} = \half F_{MNRS}F_{PQ}{}^{RS}~.
\end{equation*}
Tracing the above equation in $N,Q$ we recover the Einstein field
equations
\begin{equation}
  R_{MN} = \half F^2_{MN} - \tfrac1{6} g_{MN} F^2 = 0~,
\end{equation}
and in particular a relation between the Ricci scalar and the norm of
the four-form:
\begin{equation}
  \label{eq:RF}
  R = \tfrac16 F^2~.
\end{equation}
It is clear from \eqref{eq:quaqua} that if $F=0$ the curvature of the
spacetime vanishes.  Thus the only such solution is locally isometric
to Minkowski spacetime.

\subsubsection{Terms cubic in $\Gamma$}

The vanishing of the component of $\eR$ cubic in $\Gamma$ is
\begin{multline}
  \label{eq:cubg}
  \left(\nabla_M F_{NL_1L_2L_3}-\nabla_NF_{ML_1L_2L_3}\right)\\
  -\tfrac{1}{816} \left(F_{M_1\dots M_4} F_{N_1\dots N_3N}
  \epsilon_M{}^{M_1\dots M_4N_1\dots N_3}{}_{L_1\dots L_3}
  - (N \leftrightarrow M)\right)=0~.
\end{multline}
This equation can be simplified using equation \eqref{eq:innerp}.
Indeed, in components \eqref{eq:innerp} can be written as
\begin{equation}
  F_{M[L_1L_2L_3} F_{L_4\dots L_7]}=0~.
\end{equation}
Substituting this equation back into \eqref{eq:cubg}, we find that
\begin{equation}
  \nabla_M F_{NL_1L_2L_3}- \nabla_NF_{ML_1L_2L_3}=0~.
\end{equation}
This together with the fact that $F$ is a closed four-form imply it is
parallel:
\begin{equation}
  \label{eq:paral}
  \nabla F=0~.
\end{equation}
Observe that the above equation and \eqref{eq:ling} imply the field
equations \eqref{eq:ELexplicit} of the four-form field strength.

Equation \eqref{eq:quaqua} expresses the Riemann curvature tensor
algebraically in terms of $F$ and $g$, both of which are parallel with
respect to the Levi-Civita connection.  This means that the Riemann
curvature tensor is also parallel, and we conclude that a
maximally supersymmetric solution of eleven-dimensional supergravity
is locally symmetric.  Moreover equation \eqref{eq:paral} says that
the four-form $F$ is invariant.

\subsubsection{Terms quartic in $\Gamma$}

The vanishing of the component of $\eR$ fourth order $\Gamma$ is
\begin{equation}
  \label{eq:foa}
  2 F_{L[M_1M_2M_3} F^{MNL}{}_{M_4]}-3F_{LP[M_1M_2}
  F^{LP}{}_{M_3}{}^{[N} \delta^{M]}{}_{M_4]}=0~.
\end{equation}
Antisymmetrising in all free indices, we find
\begin{equation}
  F_{L[M_1M_2M_3} F^L{}_{M_4M_5M_6]}=0~.
\end{equation}
Antisymmetrising in five of the six free indices in \eqref{eq:foa}, we
find
\begin{equation}
\label{eq:foc}
  2F_{L[M_1M_2M_3} F^{LM}{}_{M_4M_5]}-\tfrac{3}{2} F_{LP[M_1M_2}
  F^{LP}{}_{M_3M_4} \delta^M{}_{M_5]}=0~.
\end{equation}

Next we contract the indices $M$ and $M_4$ in \eqref{eq:foa} to find
\begin{equation}
  F_{LP[M_1M_2} F^{LP}{}_{M_3]}{}^N=0~.
\end{equation}
This in turn implies that
\begin{equation}
  \label{eq:fod}
  F_{LP[M_1M_2} F^{LP}{}_{M_3M_4]}=0~.
\end{equation}

\subsubsection{Terms quintic in $\Gamma$}

Let us next investigate the conditions that arise from the vanishing
of the fifth order terms in $\Gamma$.  These are
\begin{multline}
  \label{eq:fioa}
  -2 F^M{}_{[P_1P_2P_3} F^N{}_{Q_1Q_2Q_3]}-2 F^M{}_{L[P_2P_3}
 \delta^N{}_{P_1} F^L{}_{Q_1Q_2Q_3]}\\
 +2 F^N{}_{L[P_2P_3} \delta^M{}_{P_1} F^L{}_{Q_1Q_2Q_3]} + 3
 \delta^{MN}{}_{[P_1P_2} F_{|LP| P_3Q_1} F^{LP}{}_{Q_2Q_3]}=0~,
\end{multline}
where $\delta^{MN}_{PQ}=\delta^{[M}_P \delta^{N]}_Q$.  Combining
\eqref{eq:foa}, \eqref{eq:fod} with \eqref{eq:fioa}, we find that
\begin{equation}
  \label{eq:fiob}
  F_{M[P_1P_2P_3} F_{Q_1Q_2Q_3]N}=0
\end{equation}
or equivalently
\begin{equation}
  \label{eq:fioc}
  \iota_X F \wedge \iota_Y F = 0~.
\end{equation}
This concludes the investigation of the various conditions that arise
from the vanishing of the curvature of the supercovariant connection.

\subsection{$F$ is decomposable }
\label{sec:decomp}

To analyse further the conditions that we have derived in the previous
section, we shall use the \emph{Plücker relations}.  A $p$-form is
said to be \emph{decomposable} if it can be written as the wedge
product of $p$ one-forms.  It is a classical result in algebraic
geometry (see, for example, \cite[Chapter~1]{GH}) that a $p$-form $F$
is decomposable if and only if
\begin{equation}
  \label{eq:pluecker}
  \iota_\Xi\, F \wedge F = 0
\end{equation}
for every $(p-1)$-multivector $\Xi$.  In this section we shall show
that the conditions \eqref{eq:innerp} and \eqref{eq:fioc} derived in
the previous section actually imply \eqref{eq:pluecker} and hence that
$F$ is decomposable.  Observe that the converse is trivially true: if
$F$ is decomposable then \eqref{eq:innerp} and \eqref{eq:fioc} are
satisfied.

Our starting point is equation \eqref{eq:innerp}.  Contracting this
equation with another vector field $Y$, we obtain
\begin{equation*}
  \iota_Y\iota_X F \wedge F - \iota_X F\wedge \iota_Y
  F = \iota_Y \iota_X F \wedge F=0~,
\end{equation*}
where to establish the first equality we have used \eqref{eq:fioc}.
Contracting the above equation with another vector field $Z$, we find
\begin{equation}
  \label{eq:ffg}
  \iota_Z \iota_Y \iota_X F \wedge F = - \iota_Y \iota_X F\wedge
  \iota_Z F~.
\end{equation}
Next contracting equation \eqref{eq:fioc} with a third vector field,
we get
\begin{equation}
  \iota_Y \iota_X F \wedge \iota_Z F = \iota_X F \wedge \iota_X F
  \wedge \iota_Y \iota_Z F= \iota_Y \iota_Z F\wedge \iota_X F~.
\end{equation}
Therefore the expression in the right-hand-side of \eqref{eq:ffg} is
symmetric in $X$ and $Z$, whereas the left-hand-side of \eqref{eq:ffg}
is skew-symmetric.  This means that both terms in \eqref{eq:ffg} must
vanish separately.  In particular we find that
\begin{equation}
  \iota_Z\iota_Y\iota_X F\wedge F=0~,
\end{equation}
which is precisely equation \eqref{eq:pluecker}.

With all but equation \eqref{eq:quaqua} fully analysed we can already
conclude that a solution $(M,g,F)$ of eleven-dimensional supergravity
is maximally supersymmetric if and only if $(M,g)$ is a locally
symmetric space and $F$ is parallel and decomposable.

\subsection{The local geometries}
\label{sec:local}

In this section we narrow down the possible choices of symmetric
spaces that are maximally supersymmetric solutions of M-theory by
exploiting the information that $F$ is a parallel decomposable form in
a (lorentzian) symmetric space.  We will achieve a characterization of
the geometry up to local isometry.  First recall that if the four-form
$F$ vanishes, the only solution is Minkowski spacetime up to discrete
identifications which preserve supersymmetry.  In what follows we
assume that $F\not=0$.

We will be making use of the classification of lorentzian symmetric
spaces by Cahen and Wallach \cite{CahenWallach}.  They stated the
following theorem:

\begin{thm}
  \label{thm:CahenWallach}
  Let $(M,g)$ be a simply-connected lorentzian symmetric space.  Then
  $M$ is isometric to the product of a simply-connected riemannian
  symmetric space and one of the following:
  \begin{itemize}
  \item $\RR$ with metric $-dt^2$;
  \item the simply-connected covering space of $D$-dimensional
    (anti)~de Sitter space, where $D\geq 2$; or
  \item an Hpp-wave $\CW_D(A)$ with $D\geq 3$ and metric given by
    \eqref{eq:CWmetric}.
  \end{itemize}
\end{thm}

If we drop the hypothesis of simply-connectedness then this theorem
holds up to local isometry.  We will make use of this result
repeatedly.

Let $(M,g,F)$ be a maximally supersymmetric solution.  As we have seen
$(M,g)$ is a locally symmetric space, whence locally isometric to one
of the spaces in the list in the above theorem.  Every such space is
acted on transitively by a Lie group $G$ (the group of
\emph{transvections}), whence if we fix a point in $M$ (the
\emph{origin}) with isotropy $H$, $M$ is isomorphic to the space of
cosets $G/H$.  Let $\fg$ denote the Lie algebra of $G$ and $\fh$ the
Lie subalgebra corresponding to $H$.  Then $\fg$ admits a vector space
decomposition $\fg=\fh\oplus \fm$, where $\fm$ is isomorphic to the
tangent space of $M$ at the origin.  The Lie brackets are such that
\begin{equation*}
  [\fh,\fh] \subset \fh \qquad   [\fh,\fm] \subset \fm \qquad
  [\fm,\fm] \subset \fh~.
\end{equation*}
The metric $g$ on $M$ is determined by an $\fh$-invariant inner
product $B$ on $\fm$.  Since the four-form $F$ is parallel, it is
$G$-invariant.  This means that it is uniquely defined by its value at
the origin, which defines an $\fh$-invariant four-form on $\fm$.
Since it does not vanish (by hypothesis) and is decomposable, it
determines a four-dimensional vector subspace $\fn \subset \fm$ as
follows: if at the origin $F = \theta_1 \wedge \theta_2 \wedge
\theta_3 \wedge \theta_4$, then $\fm$ is the span of (the dual vectors
to) the $\theta_i$.  Furthermore, because $F$ is invariant, we have
that $H$ leaves the space $\fn$ invariant, whence $[\fh, \fn] \subset
\fn$, which means that the holonomy group of $M$ (which is isomorphic
to $H$) acts reducibly.  In lorentzian signature this does not imply
that the space is locally isometric to a product, since the metric
may be degenerate when restricted to $\fn$.  Therefore we must
distinguish between two cases, depending on whether or not the
restriction $B|_{\fn}$ of $B$ to $\fn$ is or is not degenerate.

If $B|_{\fn}$ is non-degenerate, then it follows from the de~Rham--Wu
decomposition theorem \cite{Wu} that the space is locally isometric to
a product $N\times P$, with $N$ and $P$ locally symmetric spaces of
dimensions four and seven, respectively.  Explicitly, we can see this
as follows: there exists a $B$-orthogonal decomposition
$\fm=\fn\oplus\fp$, with $\fp := \fm^\perp$, where $[\fh,\fp] \subset
\fp$ because of the invariance of the inner product.  Let $\fg_N = \fh
\oplus \fn$ and $\fg_P = \fh \oplus \fp$.  They are clearly both Lie
subalgebras of $\fg$.  Let $G_N$ and $G_P$ denote the respective
(connected, simply-connected) Lie groups.  Then $N$ will be locally
isometric to $G_N/H$ and $P$ will be locally isometric to $G_P/H$, and
$M$ will be locally isometric to the product.  The metrics on $N$ and
$P$ are induced by the restrictions of $\fn$ and $\fp$ respectively of
the inner product $B$ on $\fn \oplus \fp$, denoted
\begin{equation}
  \begin{aligned}
    B_\fn&= B|_\fn
    \\
    B_\fp&=B|_\fp~.
  \end{aligned}
\end{equation}
We shall denote the metrics on $N$ and $P$ induced from the above
inner products by $h$ and $m$, respectively.

On the other hand if the restriction $B|_\fn$ is degenerate, so that
$\fn$ is a null four-dimensional subspace of $\fm$, the four-form $F$
is also null.  From Theorem~\ref{thm:CahenWallach} one sees (see,
e.g., \cite{FOPflux}) that the only lorentzian symmetric spaces
admitting parallel null forms are those which are locally isometric to
a product $M=\CW_d(A)\times Q_{11-d}$, where $\CW_d(A)$ is a
$d$-dimensional Cahen-Wallach space and $Q_{11-d}$ is an
$(11{-}d)$-dimensional riemannian symmetric space.

In summary, there are two separate cases to consider:
\begin{enumerate}
\item $(M,g)= (N_4 \times P_7, h \oplus m)$ (locally), where
  $(N,h)$ and $(P,m)$ are symmetric spaces and where $F$ is
  proportional to (the pull-back of) the volume form on $(N,h)$; or
\item $M = \CW_d(A) \times Q_{11-d}$ (locally) and $d\geq 3$, where
  $Q_{11-d}$ is a riemannian symmetric space.
\end{enumerate}

The first case corresponds to the well-known Freund--Rubin Ansatz,
which we have \emph{derived} here from the requirement of maximal
supersymmetry.  The second case has also been considered before
\cite{KG,FOPflux}.  The Cahen--Wallach metrics are special cases of
metrics admitting parallel null spinors
\cite{Bryant-ricciflat,JMWaves}.  This larger class of metrics have
appeared in the supergravity literature, see for example
\cite{MWave}.

We now investigate each of these two cases above separately.

\subsubsection{The Freund--Rubin Ansatz revisited}
\label{sec:FR}

We start by reconsidering the Freund--Rubin Ansatz: $(M,g) =
(N_4\times P_7, h \oplus m)$, locally.  Since $M$ is locally
symmetric, we can analyze the equations implied by supersymmetry at
the origin of the symmetric space. It is straightforward to see that
the only non-trivial equation that remains to be solved is
\eqref{eq:quaqua}.  Since the spacetime is isomorphic to a product,
the curvature of spacetime decomposes into the curvatures of $N$ and
$P$.

Since $F$ vanishes along $P$, the curvature of $P$ is
\begin{equation}
  R_{abcd}=-\tfrac{1}{3} R \left(m_{ac} m_{bd}-m_{ad} m_{bc}\right)~,
\end{equation}
where $a,b,c,d=1,2, \dots, 7$ label the coordinates of $P$ and $R$ is
the Ricci scalar of $M$.   This shows that $P$ is a space form.  If
$R<0$ then $P$ is locally isometric to $S^7$ with Ricci scalar $R_P
=-7R$, whereas if $R>0$, $P$ is isometric to $\AdS_7$ with Ricci scalar
$-7R$.

Similarly the curvature of $N$ is obtained by evaluating equation
\eqref{eq:quaqua} along the directions of $\fn$.  After some 
computation, we find that
\begin{equation}
  R_{ijkl}=-\tfrac{2}{3} R \left(h_{ik} h_{jl}-h_{il} h_{jk}\right)~.
\end{equation}
Thus if $R<0$, then $N$ is locally isometric to $\AdS_4$ with Ricci
scalar $R_N = 8R$, whereas if $R>0$, then $N$ is locally isometric to
$S^4$ with Ricci scalar $R_N = 8R$.  Notice that $R_P + R_N = R$, as
it should since $R$ is the scalar curvature of $M=P\times Q$.

\subsubsection{The case of a null four-form}
\label{sec:null}

If $F^2=0$, we have shown that $M = \CW_d(A) \times Q_{11-d}$, for
$3\leq d \leq 11$, where $Q$ is a riemannian symmetric space.
Since $F$ is decomposable and null, it must be of the form
\begin{equation}
  \label{eq:nff}
  F= dx^- \wedge \varphi~,
\end{equation}
where $dx^-$ is (up to scale) a parallel null $1$-form, which exists
in every $\CW_d(A)$, and $\varphi$ is a parallel three-form on $M$
with positive norm: $\varphi^2>0$.

Substituting \eqref{eq:nff} into the expression of the curvature
tensor \eqref{eq:quaqua}, we find that the curvature of $Q$ vanishes.
Therefore, $M$ is locally isometric to $\CW_d(A) \times \RR^{11-d}$.

The metric on $M =\CW_d(A) \times \RR^{11-d}$ can be written in local
coordinates as follows
\begin{equation*}
  ds^2 = 2 dx^+ dx^- + \sum_{i,j=1}^9 A_{ij} x^i x^j (dx^-)^2 +
  \sum_{i=1}^9 (dx^i)^2~.
\end{equation*}
where $A$ is a symmetric $9 \times 9$ matrix which is degenerate along
the $\RR^{11-d}$ directions.  In addition, we can always choose
coordinates in $\RR^9$ in such a way that the parallel, decomposable
$4$-form $F$ is given by
\begin{equation*}
  F = \mu\, dx^- \wedge dx^1 \wedge dx^2 \wedge dx^3~,
\end{equation*}
where $\mu$ is some constant.  As shown in \cite{FOPflux} this
is the ansatz for an Hpp-wave and as shown in that paper, the only
maximally supersymmetric solution is the one in \cite{KG}, for which
$A=-\frac{\mu^2}{36} \text{diag}(4, 4, 4, 1, \dots,1)$.

In summary, we have proven Theorem~\ref{th:d=11}, stated in the
introduction.

\subsection{Maximal supersymmetry in IIA supergravity}

Type IIA supergravity \cite{GianiPerniciIIA, CampbellWestIIA,
  HuqNamazieIIA} is obtained by dimensional reduction from
eleven-dimensional supergravity.  This means that any solution of IIA
supergravity can be uplifted (or oxidised) to a solution of
eleven-dimensional supergravity possessing a one-parameter subgroup of
the symmetry group such that reducing along its orbits yields the IIA
solution we started out with.  If the IIA supergravity solution
preserves some supersymmetry, its lift to eleven dimensions will
preserve at least the same amount of supersymmetry.  This means that a
maximally supersymmetric solution of IIA supergravity will uplift to
one of the maximally supersymmetric solutions of eleven-dimensional
supergravity determined in the previous section.  Therefore the
determination of the IIA vacua reduces to classifying those
dimensional reductions of the eleven-dimensional vacua which preserve
all supersymmetry.

As explained already in \cite{FOPflux}, the only such reductions are
the reductions of the flat eleven-dimensional vacuum by a translation
subgroup of the Poincaré group. This can also be verified by an
explicit analysis of the Killing spinor equations of IIA supergravity.

In summary, this proves Theorem~\ref{th:IIA}, stated in the
introduction.

\section{Maximal supersymmetry in IIB supergravity}
\label{sec:IIB}

The bosonic field content of IIB supergravity \cite{SchwarzIIB,
  SchwarzWestIIB, HoweWestIIB} is the metric $g$, two scalars, two
complex three-form gauge potentials $\{A^1, A^2: A^1=(A^2)^*\}$ and a
real four-form gauge potential $A$. The two scalars parametrise the
upper half-plane $\SU(1,1)/\U(1)$, the two three-form gauge potentials
$\{A^i; i=1,2\}$ transform as a $\SU(1,1)$-doublet while $A$ is a
$\SU(1,1)$-singlet. We follow mostly the notation of \cite{SchwarzIIB}.

We shall not state the field equations of IIB supergravity here.  This
is because, as in the case of eleven-dimensional supergravity that we
have already studied, the conditions for maximal supersymmetry derived
from the Killing spinor equations imply all the field equations.  To
continue, we follow \cite{SchwarzIIB} and define the fields
\begin{equation}
  \label{eq:formdef}
  \begin{aligned}
    P_M&=-\varepsilon_{ij} V^i_+ \partial_M V^j_+ \\
    G_3&=-\varepsilon_{ij} V^i_+ F^j \\
    F_5&=dA+\tfrac{i}{ 16}\varepsilon_{ij} A^i\wedge F^j
  \end{aligned}
\end{equation}
where $\{V^i_a:i=1,2, a=+,-\}$ are $\SU(1,1)$ matrices parameterized by
the two scalars which are $\SU(1,1)$-doublets under rigid
transformations. In addition $V^i_+$ transforms with the standard
one-dimensional complex representation on $\U(1)$ under a local
transformation and $V^i_-$ transforms with its conjugate
representation; $V_+^i=(V^i_-)*$.  Moreover $F^i=dA^i$, $F$ is a
self-dual five-form and is a singlet under both rigid $\SU(1,1)$ and
local $\U(1)$ transformations, and $G$ is a singlet under rigid
$\SU(1,1)$ tranformations but transforms under the local $\U(1)$
transformations.

Next introduce the canonical $\U(1)$ connection
\begin{equation}
  Q=-i\varepsilon_{ij} V^i_- d V^j_+
\end{equation}
on the coset space $\SU(1,1)/\U(1)$.  The Killing spinor equations of
IIB supergravity are given by
\begin{equation}
  \label{eq:iibk}
  \begin{split}
    \eD_M\varepsilon &= D_M\varepsilon+ \tfrac{i}{192} F_{L_1\dots L_5}
    \Gamma^{L_1\dots L_5}\Gamma_M\varepsilon \\
    &\qquad + \tfrac{1}{ 96} \big(\Gamma_M{}^{L_1L_2L_3} \hat
    G_{L_1L_2L_3} - 9 \Gamma^{L_1L_2}  G_{ML_1L_2}\big)\varepsilon^*=0\\
    \Gamma^M\varepsilon^* & P_M-\tfrac{1}{24} \Gamma^{MNR} G_{MNR}
    \varepsilon = 0~,
  \end{split}
\end{equation}
where
\begin{equation}
  D_M=\nabla_M-\tfrac{i}{2}Q_M~,
\end{equation}
$Q_M$ is the pull-back of the connection of $\U(1)$ connection of the
coset on to the spacetime, $\varepsilon$ is a complex Weyl spinor,
$\Gamma^{11}\varepsilon=\varepsilon$.

There are two types of Killing spinor equations for IIB supergravity.
One is a parallel transport type of equation similar to that we have
investigated in the context of eleven-dimensional supergravity. The
other is an algebraic equation which does not involve derivatives on
the spinor $\varepsilon$. Since we are seeking maximally supersymmetric
solutions, the components of the algebraic Killing spinor equation as
expanded in a basis of the Clifford algebra should vanish. This in
particular implies that
\begin{equation}
  \label{eq:algcon}
  P_M=0 \qquad\text{and}\qquad G_{MNR}=0~.
\end{equation}

Substituting the second equation above into the supercovariant
derivative in \eqref{eq:iibk}, we find that it simplifies to
\begin{equation}
\eD_M\varepsilon=D_M\varepsilon+ \tfrac{i}{192} F_{L_1\dots L_5}
 \Gamma^{L_1\dots L_5}\Gamma_M\varepsilon~.
\end{equation}

The strategy that we shall adopt to find the maximally supersymmetric
solutions is to compute the curvature $\eR$ of the supercovariant
derivative $\eD$ above as we have done for eleven-dimensional
supergravity. Indeed after some computation, we find that
\begin{multline}
   \label{eq:iibcurv}
   \eR_{MN}={\eF}_{MN}+\tfrac{1}{4} R_{MNPQ} \Gamma^{PQ}+\tfrac{i}{192}
   \nabla_MF_{NPQRS} \Gamma^{PQRS}\\
   - \tfrac{i}{192} \nabla_NF_{MPQRS}\Gamma^{PQRS}
   - \tfrac1{192} F_{ML_1L_2L_3 P} F_N{}^{L_1L_2L_3}{}_{Q}
   \Gamma^{PQ}\\
   + \tfrac1{12\cdot 64^2} F_{LMM_1M_2M_3} F^L{}_{NN_1N_2N_3}
     \varepsilon^{M_1M_2M_3N_1N_2N_3}{}_{PQRS} \Gamma^{PQRS}~,
\end{multline}
where ${\eF}_{MN}=-\frac{i}{2} (\partial_M Q_N-\partial_N Q_M)$.

\subsection{The vanishing of the algebraic Killing spinor equation conditions}

Here we shall show that the conditions \eqref{eq:algcon} of the
algebraic Killing spinor equation required by maximal supersymmetry
imply that the two scalars are constant and that $F^i=0$. This is most
easily seen by fixing the local $\U(1)$ symmetry of the coset space.
This has been done in \cite{SchwarzIIB} and so we shall not repeat the
computation here. After gauge fixing, the theory has two real scalars
parameterized by the complex scalar $B$ and a complex three three form
field strength $F$.  The final expressions for the relevant fields are
the following:
\begin{equation}
  \label{eq:gauga}
  \begin{split}
    G_{MNR}&= f (F_{MNR}-B F^*_{MNR})\\
    P_M&=f^2 \partial_MB
  \end{split}
\end{equation}
where $f^{-2}=1-BB^*$, $BB^*<1$.  The condition $P_M=0$ in
\eqref{eq:algcon} implies that the complex scalar field $B$ is
constant. The vanishing of $G=0$ in \eqref{eq:algcon} implies that
$F_{MNR}=0$. So the only fields that remain to be determined by
maximal supersymmetry are the metric $g$ and the self-dual five-form
field strength $F_5$.  In addition from \eqref{eq:formdef} and
\eqref{eq:gauga}, the five-form self-dual field strength is $F_5=dA$.
In particular, $F_5$ is closed and since it is self-dual, it is also
co-closed.

Another consequence of the algebraic Killing spinor equation is that
the pull-back of the $\U(1)$ curvature of the coset space
$\SU(1,1)/\U(1)$ on the spacetime vanishes. This can be seen by the
formula similar to those in \eqref{eq:gauga} which expresses the
pull-back $\U(1)$ connection in terms of $B$ as
\begin{equation}
Q_M=f^2 {\rm Im}\big( B\partial_M B^*\big)
\end{equation}
Since $B$ is constant, $Q_M=0$.

\subsection{The vanishing of curvature conditions}

\subsubsection{Terms zeroth order in $\Gamma$}

This term involves the pull-back of the curvature of the $\U(1)$
connection of coset space on the spacetime, ie
\begin{equation}
{\eF}_{MN}=0~.
\end{equation}
This however vanishes as a consequence of the $Q_M=0$ condition derived
in the previous section from the algebraic Killing spinor
equations. So there is no additional condition.

\subsubsection{Terms quadratic in $\Gamma$}

The condition on the quadratic terms in $\Gamma$ is the following:
\begin{equation}
  \tfrac{1}{4} R_{MNPQ}-\tfrac{1}{192} F_{ML_1L_2L_3[P}
  F_{|N|}{}^{L_1L_2L_3}{}_{Q]}=0 
\label{eq:qoiib}
\end{equation}
Contracting $N$ and $Q$, we find the Einstein field equations
\begin{equation}
  R_{MN}= \tfrac{1}{96} F_{ML_1L_2L_3L_4} F_N{}^{L_1L_2L_3L_4}~.
\end{equation}
In particular the Ricci scalar vanishes, $R=0$, because $F$ is
self-dual.

\subsubsection{Terms fourth order in $\Gamma$}

The condition on the fourth order terms in $\Gamma$ is the following:
\begin{multline}
    \tfrac{i}{192}\nabla_MF_{NPQRS}-\tfrac{i}{192} \nabla_NF_{MPQRS}
    \\
    + \tfrac1{12 \cdot 64^2} F_{LMM_1M_2M_3} F^L{}_{NN_1N_2N_3}
    \varepsilon^{M_1M_2M_3N_1N_2N_3}{}_{PQRS}=0
\end{multline}
In particular the imaginary and real parts of the above equation
should vanish separately. Since $F$ is real, this implies that
\begin{equation}
  \label{eq:fobiib}
  \nabla_MF_{NPQRS}- \nabla_NF_{MPQRS}=0
\end{equation}
and
\begin{equation}
  \label{eq:fociib}
  F_{LMM_1M_2M_3} F^L{}_{NN_1N_2N_3}
  \varepsilon^{M_1M_2M_3N_1N_2N_3}{}_{PQRS}=0~.
\end{equation}
Antisymmetrising \eqref{eq:fobiib} in the indices $M,N,P,Q,R$  and using
that $dF=0$, we find that $F$ is
parallel with respect to the Levi-Civita connection.  Combining this
fact with equation \eqref{eq:qoiib} one concludes that the curvature
tensor is parallel, whence $(M,g)$ is locally symmetric.

The equation \eqref{eq:fociib} can be rewritten in a more invariant
form as
\begin{equation}
  \label{eq:fodiib}
  \iota_X F_L\wedge \iota_Y F^L=0~.
\end{equation}

Next contracting $M$ with $P$ in \eqref{eq:fociib}, the resulting
equation can be written as
\begin{equation}
  \iota_XF_L\wedge F^L=0
\end{equation}
Taking the inner derivation of this equation with respect to the vector
field $Y$ and using the equation \eqref{eq:fodiib}, we find that
\begin{equation}
  \iota_Y\iota_X F_L\wedge F^L=0~.
\end{equation}

Now take the inner derivation of this equation with another vector
field $Z$. This gives
\begin{equation}
  \label{eq:fooiib}
  \iota_Z\iota_Y\iota_X F_L\wedge F^L+\iota_Y\iota_X F_L\wedge \iota_Z
  F^L = 0~.
\end{equation}
Taking the inner derivation of $\iota_X F_L\wedge \iota_Z F^L=0$ with
respect to $Y$, we find
\begin{equation}
  \iota_Y\iota_X F_L\wedge \iota_ZF^L=\iota_Y\iota_Z F_L\wedge \iota_X
  F^L
\end{equation}
which implies that the right hand side of \eqref{eq:fooiib} is
symmetric in the interchange of $X$ and $Y$ while the left hand side
is skew-symmetric. This implies that both terms in \eqref{eq:fooiib}
should vanish separately.  In particular, we have that
\begin{equation}
\label{eq:threein}
  \iota_Z\iota_Y\iota_X F_L\wedge F^L=0~.
\end{equation}

This condition \eqref{eq:threein} is analogous to the Plücker relations
which appear in eleven-dimensional supergravity.  It is therefore
conceivable that it should imply a decomposition of the self-dual
five-form $F$.  In fact, it can be shown that equation
\eqref{eq:threein} implies that
\begin{equation}
  \label{eq:decomp}
  F = G + \star\, G~,
\end{equation}
where $G$ is a decomposable five-form.  This is proven in
\cite{FOPPluecker}, where we also state a conjectural generalisation
involving $p$-forms in euclidean or lorentzian $n$-dimensional vector
spaces, which we verify in low dimension.  Notice that since $F$ is
parallel, then so is $G$.

\subsection{The local geometries}

We must distinguish between two cases, depending on whether or not the
five-form $G$ in equation \eqref{eq:decomp} is null.  First suppose
that $G$ (and hence $F$) is not null.  An analysis similar to that
presented for eleven-dimensional supergravity shows that the five-form
$G$ induces a local decomposition of $(M,g)$ into a product $N_5\times
P_5$ of two five-dimensional symmetric spaces $(N,h)$ and $(P,m)$,
where $G \propto \dvol(N)$ and hence $\star G \propto \dvol(P)$.  Thus
again we have derived the Freund--Rubin ansatz from maximal
supersymmetry.  Since $(M,g)$ is lorentzian, one of the spaces $(N,h)$
and $(P,m)$ is lorentzian and the other riemannian.  If the norm of
$G$ is positive, then $N$ is riemannian and $P$ is lorentzian, and
vice-versa if $G$ is negative.  By interchanging $G$ with $\star G$ if
necessary, we can assume that $G$ has positive norm and hence that $N$
is riemannian.

We continue as in the eleven-dimensional case by analysing the
remaining condition \eqref{eq:qoiib}.  Evaluating this along the
directions of $N_5$, we find that
\begin{equation}
  R_{ijkl}= \tfrac1{16} G^2 \left(h_{ik} h_{jl}-h_{jk} h_{il}\right)~,
\end{equation}
whence $N$ is locally isometric to $S^5$.  Similarly, evaluating
equation \eqref{eq:qoiib} along $P$, we find that
\begin{equation}
  R_{abcd}= - \tfrac1{16} G^2 \left(m_{ac} m_{bd}-m_{bc}
    m_{ad}\right)~,
\end{equation}
whence $P$ is locally isometric to $\AdS^5$.  Both $\AdS_5$ and $S^5$
have the same radii of curvature which are related to the norm of $G$
in the following way: if $S^5$ has Ricci scalar $R$, then $G =
2 \sqrt{\tfrac{R}5} \dvol(S^5)$.

Next suppose that $G$ is null.  As in the discussion of the
eleven-dimensional case, and using the fact that the only symmetric
spaces with null parallel forms are the CW Hpp-waves, the spacetime is
locally isometric to $(M,g)=(\CW_d(A)\times Q_{10-d}, h\oplus m)$,
where $d\geq 3$, and the five-form field strength is
\begin{equation}
  F= dx^- \wedge \varphi~,
\end{equation}
where $\varphi$ is a self-dual four-form and $dx^-$ is the one-form
dual to the parallel null vector in $\CW_d(A)$.  Evaluating
\eqref{eq:qoiib} along the directions of $Q_{10-d}$, we find that the
curvature of $Q_{10-d}$ vanishes because $F$ is null.  An analysis
similar to that made for the eleven-dimensional case reveals that the
metric and five-form of spacetime can be written as
\begin{equation}
  \begin{aligned}
    ds^2&=2dx^+ dx^-+\sum_{i,j}^8 A_{ij} x^i x^j (dx^-)^2+
    \sum_{i=1}^8 (dx^i)^2\\
    F_5&= \mu dx^-\wedge (dx^{1234}+dx^{5678})~.
  \end{aligned}
\end{equation}
This is precisely the ansatz used to find the maximally supersymmetric
Hpp-wave solution of IIB supergravity \cite{NewIIB}.  Here we have
derived it from the requirement of maximal supersymmetry.  In
particular, the maximally supersymmetric Hpp-wave solution occurs for
$A=-\tfrac{\mu^2}{16} {\rm diag}(1,1,\dots,1)$.

This concludes the proof of Theorem~\ref{th:IIB}, stated in the
introduction.

\section{Maximal supersymmetry in other ten-dimensional supergravities}
\label{sec:other}

In this section we discuss the remaining ten-dimensional supergravity
theories: the heterotic, type I and massive IIA supergravities.

\subsection{Heterotic supergravities}

The only maximally supersymmetric solution of heterotic supergravities
is the ten-dimensional Minkowski spacetime with constant dilaton and
rest of form-field strengths to vanish.  This can be easily seen by
inspecting the Killing spinor equations
\begin{equation}
  \begin{aligned}
    \nabla^+\varepsilon&=0\\
    \Gamma^M\partial_M\phi\varepsilon-\tfrac{1}{12} H_{MNR}
    \Gamma^{MNR}\varepsilon&=0\\
    F_{MN}\Gamma^{MN}\varepsilon&=0~,
  \end{aligned}
\end{equation}
where $\nabla^+ = \nabla + \tfrac{1}{2} H$, $H$ is the NSNS three-form
field strength, $\phi$ is the dilaton, $\varepsilon$ is a
Majorana-Weyl sixteen-component spinor and $F$ is the curvature of the
gauge connection with gauge group $E_8\times E_8$ or
$\Spin(32)/\ZZ_2$.  (We have suppressed gauge indices).  From the
second Killing spinor equation one concludes that $\phi$ is constant
and $H=0$.  The first equation implies that the curvature of the
Levi-Civita connection vanishes and so the spacetime is locally
isometric to Minkowski spacetime.  The last Killing spinor equation
implies that the curvature of the gauge fields vanishes and so for
simply connected spacetimes the gauge connection vanishes as well.
This proves the second part of Theorem~\ref{th:other} in the
introduction.

\subsection{Type I supergravity}

Similarly, the only maximally supersymmetric solution of type I
supergravity is Minkowski spacetime. This again can be easily seen by
inspecting the Killing spinor equations
\begin{equation}
  \begin{aligned}
    \nabla_M\eta + \tfrac{1}{8} H_{MNR} \Gamma^{NR}\eta&=0 \\
    \Gamma^M\partial_M\phi\eta + \tfrac{1}{12} H_{MNR}\Gamma^{MNR}\eta&=0 \\
    F_{MN}\Gamma^{MN}\eta&=0~,
  \end{aligned}
\end{equation}
where $H$ is the RR three-form field strength, $\phi$ is the dilaton,
$\varepsilon$ is a Majorana-Weyl sixteen component spinor and $F$ is
the curvature of the gauge connection with gauge group
$\Spin(32)/\ZZ_2$. (We have suppressed gauge indices). Again the
second Killing spinor equation implies that $\phi$ is constant and
$H=0$.  Then the first implies that the curvature of the Levi-Civita
connection vanishes and so the spacetime is locally isometric to
Minkowski space.  The last Killing spinor equation implies that the
curvature of the gauge fields vanishes and so for simply connected
spacetimes the gauge connection vanishes as well.  This proves the
first part of Theorem~\ref{th:other} in the introduction.

\subsection{Massive IIA supergravity}

Although Romans' massive IIA supergravity \cite{Romans-massive} does
admit supersymmetric solutions, e.g., the D8-brane
\cite{PolchinskiWitten, BdRGPT}, it has no vacua for nonzero mass
parameter.  This can be easily seen by investigating the dilatino
Killing spinor equation of the theory:
\begin{multline}
  \left( \partial_M\phi \Gamma^M + \tfrac{5}{4} m e^{\frac{5}{4}\phi}
    - \tfrac{3}{4} m e^{\frac{3}{4}\phi} B_{MN}
    \Gamma^{MN}\Gamma_{11}\right. \\
  - \left. \tfrac{1}{6} e^{-\frac12 \phi} H_{MNR}
    \Gamma^{MNR}\Gamma_{11} + \tfrac{1}{48} e^{\frac{1}{4}\phi}
    F_{MNRP} \Gamma^{MNRP}\right)\varepsilon=0~,
\end{multline}
where $B$ is the two-form gauge potential, $H=dB$, $F=dC+m B\wedge B$,
$\phi$ is the dilaton and $m$ is the cosmological constant.  For
maximal supersymmetry every term in the above Killing spinor equation
must vanish separately. The first term implies that $\phi$ is
constant, the third term implies that $B=0$ and so $H=0$, the last
term implies that $F=0$. However the second term cannot be made to
vanish because $m$ is nonzero. In summary, this proves
Theorem~\ref{th:massive} in the introduction.  Note that it is not
straightforward to take the limit $m\rightarrow 0$ in massive IIA to
recover the usual IIA supergravity.  This can only be done after
appropriate redefinitions of the fields.

\section*{Acknowledgments}

This work was completed while JMF was visiting the IHÉS and it is his
pleasure to thank them and in particular Jean-Pierre Bourguignon for
the invitation, their support and for providing an ideal environment
in which to do research.

JMF is a member of EDGE, Research Training Network HPRN-CT-2000-00101,
supported by The European Human Potential Programme, and his research
is partially supported by the EPSRC grant GR/R62694/01.

The research of GP is partially supported by the PPARC grants
PPA/G/S/1998/00613 and PPA/G/O/2000/00451 and by the European grant
HPRN-2000-00122.

\bibliographystyle{utphys}
\bibliography{AdS,ESYM,Sugra,Geometry,CaliGeo,AdS3}

\end{document}